\def \SAIT #1 #2 {{\em Mem.\ Soc.\ Astron.\ It.\/} {\bf #1}, #2}
\def \MESS #1 #2 {{\em The Messenger\/} {\bf #1}, #2}
\def \ASTRNACH #1 #2 {{\em Astron. Nach.\/} {\bf #1}, #2}
\def \AAP #1 #2 {{\em Astron. Astrophys.\/} {\bf #1}, #2}
\def \AAL #1 #2 {{\em Astron. Astrophys. Lett.\/} {\bf #1}, L#2}
\def \AAR #1 #2 {{\em Astron. Astrophys. Rev.\/} {\bf #1}, #2}
\def \AAS #1 #2 {{\em Astron. Astrophys. Suppl. Ser.\/} {\bf #1}, #2}
\def \AJ #1 #2 {{\em Astron. J.\/} {\bf #1}, #2}
\def \ANNREV #1 #2 {{\em Ann. Rev. Astron. Astrophys.\/} {\bf #1}, #2}
\def \APJ #1 #2 {{\em Astrophys. J.\/} {\bf #1}, #2}
\def \APJL #1 #2 {{\em Astrophys. J. Lett.\/} {\bf #1}, L#2}
\def \APJS #1 #2 {{\em Astrophys. J. Suppl.\/} {\bf #1}, #2}
\def \APSS #1 #2 {{\em Astrophys. Space Sci.\/} {\bf #1}, #2}
\def \ASR #1 #2 {{\em Adv. Space Res.\/} {\bf #1}, #2}
\def \BAIC #1 #2 {{\em Bull. Astron. Inst. Czechosl.\/} {\bf #1}, #2}
\def \JSQRT #1 #2 {{\em J. Quant. Spectrosc. Radiat. Transfer\/} {\bf #1}, #2}
\def \MN #1 #2 {{\em Mon. Not. R. Astr. Soc.\/} {\bf #1}, #2}
\def \MEM #1 #2 {{\em Mem. R. Astr. Soc.\/} {\bf #1}, #2}
\def \PLR #1 #2 {{\em Phys. Lett. Rev.\/} {\bf #1}, #2}
\def \PASJ #1 #2 {{\em Publ. Astron. Soc. Japan\/} {\bf #1}, #2}
\def \PASP #1 #2 {{\em Publ. Astr. Soc. Pacific\/} {\bf #1}, #2}
\def \NAT #1 #2 {{\em Nature\/} {\bf #1}, #2}

\documentstyle{memsait}
\input epsf.sty
\begin{opening}
\title{GALACTIC MORPHOLOGIES IN DISTANT CLUSTERS} 
\author{B.M. Poggianti$^1$}
\institute{$^1$Osservatorio Astronomico, Padova, Italy}
\date{} 
\end{opening}

\begin{document}

\oddpagefooter{}{}{} 
\evenpagefooter{}{}{} 
\ 
\bigskip

\begin{abstract}
The abundance of elliptical, S0 and spiral galaxies in clusters
is presented as a function of redshift comparing the results of
the HST distant cluster studies with new data at $z\sim 0.2$ and
with the standard local comparison samples. 
This analysis confirms and strengthens the conclusion that 
the proportion of S0 galaxies increases 
toward lower redshifts at the expense of
the spiral population. The dependence of the relative occurrence
of S0s and ellipticals and of the 
morphology-density relation on the cluster concentration are also discussed.
\end{abstract}

\section{Introduction}
It is well known that cluster galaxies are systematically different
from galaxies in the field, but when and why such environmental
differences were established is still a matter of intense investigation.
Observations of distant clusters have proven
that important changes in cluster galaxies have taken place 
at a relatively recent cosmological epoch between z$\ge 0.2$ and z=0.
In this contribution I present the observational evidence for the
recent evolution of the {\em morphological types} of galaxies 
in clusters. What follows is a partial summary of two  
collaborations: with the MORPHS group (H. Butcher, A. Dressler, W. Couch,
R. Ellis, A. Oemler, I. Smail) we have studied 10 clusters at
$0.37<z<0.56$ obtaining HST images and ground based spectra, and with
G. Fasano and collaborators (W. Couch, D. Bettoni, P. Kj\ae rgaard,
M. Moles) we have analyzed the morphologies of
galaxies in 9 clusters at $0.1<z<0.25$.
This second work has made use of ground-based images taken under conditions
of very good seeing and covers a redshift range previously unexplored.
In both of these studies we are considering galaxies
down to the same absolute magnitude ($M_V$=-20) within the
central $\rm Mpc^2$ of the cluster. After carefully testing the 
accuracy of the morphologies and the consistency of the classification 
schemes in the two studies, 
we are confident we are able to broadly classify galaxies into three main
morphological types (ellipticals, S0s and spirals) and 
that the two datasets can be compared to study morphological evolution.

\section{Evolution of the fractions of morphological types}
The frequency of the various Hubble types in clusters evolve with
redshift. The different proportions of the galaxy types
at z=0.4-0.5 (Dressler et al. 1997, D97) and at z=0 (Dressler 1980, D80)
can be schematically summarized as follows:

\begin{tabular}{lll}
 &  z=0 & z=0.5 \\
 &  (D80) & (D97) \\
Sp & 1 & 2 \\
S0 & 2 & 1 \\
E  & 1 & 1($>$) \\
\end{tabular}

\vspace{0.5in}
Thus, spirals are a factor of 2 more abundant at $z\sim 0.5$ than in the 
nearby universe and S0 galaxies are proportionally less abundant, 
while the fraction of ellipticals is already as large or larger
at higher redshift. Figure~1 presents the E, S0 and spiral fraction
for clusters at low, moderate and intermediate redshift
and shows a clear evolutionary trend, with the S0 population growing
towards lower redshifts at the expense of the spiral population.
A large number of the S0s in clusters seem to be produced
from the morphological conversion of the spirals.

Let us now consider the {\em relative frequency} of S0s and ellipticals,
i.e. the number ratio of these two types, which again
shows a trend with redshift (Fig.~2). In this figure we observe a
dichotomy in the range $0.1<z<0.25$, with a group of 5 clusters having 
an S0/E ratio $\sim 1.5$ and another group of 6 clusters with a 
ratio $\sim 0.8$.
Searching for correlations between the S0/E ratio and the global cluster
properties, we found that the only structural difference
between the low-S0/E and the high-S0/E clusters is (respectively)
the presence/absence
of a high concentration of elliptical galaxies toward the cluster centre.
Hence, besides and on top of the evolutionary effects (which seem to be 
driving the
gross differences over the observed redshift range), the S0/E ratio appears
to depend also on the cluster ``type'' (its central concentration). 
A correlation between the galactic morphological content and 
the cluster concentration is known to exist at lower z, 
as discussed in Oemler (1974) (see Figs.~1 and 2). We now found that the 
dependence
of the morphological mix on the cluster type is significant at least up
to z=0.25, but we note that such a dependence appears to be smaller than the
evolutionary effects between z=0.5 and now.

\begin{figure}
\vspace{-1in}
\epsfxsize=10cm 
\hspace{3.5cm}\epsfbox{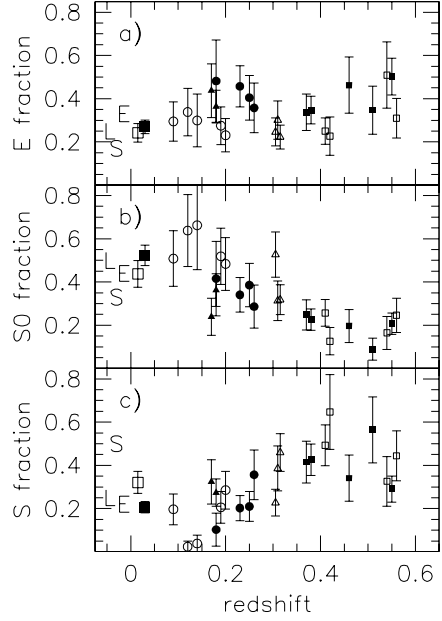} 
\vspace{-0.3in}
\caption[h]{Morphological fractions in clusters 
as a function of redshift, taken
from Fasano et al. (2000). 
Solid and open symbols indicate highly concentrated and less concentrated
clusters, respectively. The clusters 
from the sample of Fasano et al. (2000) are indicated by circles, whereas those
from the MORPHS (Smail et al. 1997)
and C98+ (Couch et al. 1998 + 3 HST archive
clusters) samples are indicated with small squares
and triangles, respectively. All these data are corrected for field 
contamination and the errorbars correspond to Poissonian values.
The average values derived for high- (large solid squares)
and low--concentration (large open squares) clusters at low-z (D80) 
are plotted at $z \sim 0$. 
Also shown are the average values from
Oemler's (1974) for different types of nearby clusters (E = elliptical-rich;
L = S0-rich; S = spiral-rich); these are placed at $z<0$ for display purposes.}
\end{figure}

\begin{figure}
\vspace{-1in}
\epsfxsize=12cm 
\hspace{3.5cm}\epsfbox{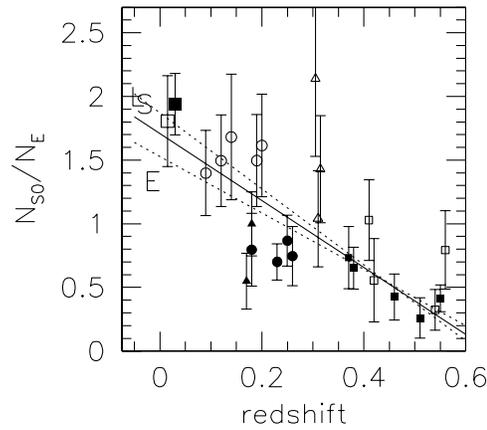} 
\vspace{-1.5in}
\caption[h]{The S0/E number ratio as a function of redshift, taken
from Fasano et al. (2000). The meaning of
the symbols is as in Fig.~1.  The linear regression of the
MORPHS data given by D97 and its 1 $\sigma$ error
are represented by the solid and dotted lines.}
\end{figure}

\section{The morphology-density relation}
In low-z clusters, Dressler (1980) found that the frequency
of each morphological type is a function of the local projected
galaxy density: ellipticals and spirals are more abundant in the 
highest and lowest density regions, respectively. 
While this relation is observed in {\em all types of clusters}
at low-z, both at z=0.5 (Dressler et al. 1997) and at z=0.1-0.2 
(Fasano et al. 2000) such a relation is found only in centrally concentrated
clusters. If confirmed by a systematic and homogeneous study of larger
galaxy samples, these results would imply that the morphology-density
relation in low-concentration clusters was established only during the
last 1-2 Gyr.



\end{document}